\documentclass[12pt]{article}

\usepackage[utf8]{inputenc}
\usepackage{graphicx}
\usepackage{amssymb}

\usepackage{natbib}
\usepackage{endnotes}
\let\footnote=\endnote

\usepackage{hyperref}

\usepackage[letterpaper]{geometry}
\begin{document}

\title{Physical, Empirical, and Conditional Inductive Possibility}
\author{{Bal\'azs Gyenis} \\ \\ Institute of Philosophy, HUN-REN RCH \\ email: {gyepi@hps.elte.hu}}
\date{}

\maketitle

\begin{abstract}
\noindent I argue that John Norton's notions of empirical, hypothetical, and counterfactual possibility can be successfully used to analyze counterintuitive examples of physical possibility and align better with modal intuitions of practicing physicists. First, I clarify the relationship between Norton's possibility notions and the received view of logical and physical possibility. In particular, I argue that Norton's empirical, hypothetical, and counterfactual possibility cannot coincide with the received view of physical possibility; instead, the received view of physical possibility is a special case of Norton's logical possibility. I illustrate my claims using examples from Classical Mechanics, General Relativity, and Quantum Mechanics. I then arrive at my conclusions by subsuming Norton's empirical, hypothetical, and counterfactual possibilities under a single concept of conditional inductive possibility and by analyzing the types and degrees of strengths that can be associated with it.
\end{abstract}

\newpage

\section{Introduction}\label{section_introduction}

\noindent \citet{Norton2022} strongly criticizes modality concepts of the philosophical literature, arguing that examples of metaphysical, epistemic, physical, nomic, and conceptual possibility and necessity can either be subsumed under his concepts of logical and empirical possibility and necessity or are not cogent and cannot be responsibly relied upon.

This paper examines the relationship between Norton's concepts of logical and empirical possibility and the received view of physical possibility. Overall, the paper is supportive of Norton's analysis: the received view of physical possibility leads to a number of counterintuitive conclusions which, I claim, can be resolved by generalizing Norton's empirical, hypothetical, and counterfactual possibility as conditional inductive possibility. On the other hand, this paper grew out of a concern that the positive results of Norton's analysis may be obscured by Norton's unusual usage of terminology and claims regarding the relationship of his new possibility concepts and that of physical and nomic possibility. 

Thus I start with a conceptual clarification of the relationship between various modality concepts. \citet{Norton2022} defines what he means by logical, empirical, hypothetical and counterfactual possibility. However, instead of defining what he means by physical and nomic possibility, Norton makes certain claims about physical and nomic possibility by using his earlier defined concepts of empirical, hypothetical and counterfactual possibility. Since the explicitly stated goal of \citet{Norton2022} is to evaluate modality concepts of the philosophical literature, the reader may infer that Norton's claims pertain to physical and nomic possibility as they are traditionally understood in the literature. 

In the first part I show that this cannot be the case. Norton claims that physical possibility is empirical possibility; however, I argue that this cannot hold if physical possibility is understood according to the received view. I also argue that Norton's hypothetical or counterfactual possibility cannot coincide with the received view either.

A simple illustration of my claims is the following: the statement that the world is completely empty describes a physically possible state of affairs according to most physical theories because it is consistent with the laws of said theories, but it cannot be empirically, hypothetically, or counterfactually possible. This is so because Norton's empirical, hypothetical, and counterfactual possibilities require positive inductive support, but any statement that is positively inductively supported in Norton's sense requires material facts pertaining to the actual circumstances on which the inductive support is based, and these material facts imply that the world cannot be entirely empty. Thus, the statement that the world is completely empty provides a straightforward example of a statement describing physically possible states of affairs (in the sense of the received view) which is neither empirically, hypothetically, nor counterfactually possible (in Norton's sense).

A further terminological complication arises because Norton also uses the term logical possibility differently from its usual interpretation in the literature, employing it in a much broader sense. As I argue, this has the consequence that the received view of physical possibility is simply a special case of Norton's logical possibility. This result also clearly indicates that the received view of physical possibility cannot coincide with any notion of possibility that derives from positive inductive support, as does Norton's empirical, hypothetical, and counterfactual possibility.

After the conceptual clarification I argue that the difference between the received view of physical possibility and empirical possibility works for the benefit of Norton's analysis: so much worse for the received view. By further developing Norton's account, I argue that the concept of conditional inductive possibility, which generalizes empirical, hypothetical, and counterfactual possibility, can resolve several counterintuitive conclusions in the philosophy of physics literature regarding what is physically possible. Among my examples are the violation of causality in Classical Mechanics, G\"odel's time travel in General Relativity, and the radical freedom of acting otherwise in Quantum Mechanics.

In Section \ref{section_physicalpossibility}, I review various formulations of physical possibility that appear in the philosophical literature. Section \ref{section_threeexamples1} presents three examples in which the received view of physical possibility leads to counterintuitive conclusions. In Section \ref{section_logicalempirical},  I review Norton's material theory of induction and his concepts of logical and empirical possibility. In Section \ref{section_emporlogical}, I demonstrate that a physical possibility of the received view is not an empirical possibility, but that the received view of physical possibility is rather a special case of Norton's logical possibility. In Section \ref{section_threeexamples2}, I compare empirical possibility with the three examples. In Section \ref{section_condposs}, I introduce a generalization of Norton's hypothetical and counterfactual possibility, conditional inductive possibility, examine the relationship between conditional inductive possibility and nomic possibility, and introduce and analyze different degrees of conditional inductive possibility. Section \ref{section_threeexamples3} returns once more to the three examples, and analyzes them from the perspective of conditional inductive possibility. The last section summarizes the findings.

\section{The Received View of Physical Possibility}\label{section_physicalpossibility}

\noindent The received view of physical possibility captures the concept of possibilities permitted by the laws of nature. Various philosophers have formulated this idea in different ways: (1) some define physical possibility in terms of possible worlds \citep[13]{Earman1986}, others in terms of events or states of affairs \citep[18]{Maudlin2007}; (2) some employ modal logic \citep[6]{Bradley-Swartz1979}, while others use non-modal logic \citep[412]{Chisholm1967}; (3) some relativize the concept to the actual world \citep[174]{Carroll1994}, while others relativize it to theories \citep[569]{Gyenis2020}.\footnote{For a detailed review of the various formulations and their different philosophical consequences, see \citet{Gyenis2020} and its references.} The following formulation, which defines the physical possibility of possible worlds in a theory-relative, modal way, is frequently utilized in the practice of philosophers of physics:
\begin{itemize}
	\item[(p)] A possible world is {\em physically possible according to theory $T$} if and only if it is consistent with the physical laws of $T$.
\end{itemize}

Other typical formulations of physical possibility can be seen as special cases of (p):

(1) Using (p) we can specify what we mean by the physical possibility of events or states of affairs: events or states of affairs are physically possible according to theory $T$ if the proposition $S$ that expresses their occurrence is true in at least one world that is physically possible according to theory $T$. I will use the notation $\Diamond^p_T S$ for the statement of physical possibility of states of affairs described by $S$ according to $T$ (note that I use superscripts on modal symbols to distinguish types of modality and subscripts to relativize a modality to a set of propositions). Physical necessity is the dual concept of physical possibility: the states of affairs described by proposition $S$ are {\em physically necessary according to theory $T$ } --- denoted in this paper by $\Box^p_T S$ --- if the negation of $S$ is physically impossible according to $T$, i.e., $\Box^p_T S = \neg \Diamond^p_T \neg S$.

(2) Instead of the modal formulation (p), some authors define physical possibility in a non-modal way. For example, \citet[412]{Chisholm1967} considers a state of affairs physically possible if the proposition $S$ expressing that the state of affairs obtained is consistent with the laws of nature. In a sound and complete logic --- such as classical first-order logic --- Chisholm's non-modal formulation is a special case of the modal formulation (p) where physically possible worlds are models of the laws. (Thus, when it does not compromise clarity, I will also use the notation $\Diamond^p_T S$ for the non-modal formulation.)

(3) If the physical laws of theory $T$ coincide with the laws of the actual world, (p) captures physical possibility relative to the actual world, according to which a possible world is physically possible if and only if it is consistent with the laws of the actual world. However, (p) is more general. Although we do not know the laws of the actual world, the theory-relative formulation allows for a clear discussion of the question: if a particular theory of physics were true, what would the physical possibilities be? Studying the physical possibilities of a particular physical theory also often leads to a deeper understanding of the theory itself. Therefore, in practice philosophers of physics tend to rely on the theory-relative formulation when discussing physical possibilities.

In summary, (p) can be considered one of the most general formulations of the received view of physical possibility in the literature.\footnote{Beyond the differences discussed above, there is a formulation of the received view of physical possibility that, instead of requiring physically possible worlds to be consistent with the natural laws of the actual world, demands that they possess exactly the same natural laws as the actual world. \citet{Gyenis2020} argues that this formulation allows for one to address the first two counterintuitive examples in this paper, but I omit this discussion here.} 

Before discussing counterintuitive examples of the received view of physical possibility, I briefly mention a related concept, {\em nomic} (or nomological) possibility. Some authors treat nomic possibility as synonymous with physical possibility, while others treat it as a generalization that does not require the laws to be {\em physical} laws. The generalization also allows us to question what possibilities are consistent with a theory of the special sciences (for example, modern inorganic chemistry). Beyond this generalization, the use of the two concepts in the literature is essentially the same, as far as I am aware.

Beyond physical and/or nomic possibility, the philosophical literature on modality distinguishes numerous concepts of possibility and necessity: logical, conceptual, metaphysical, epistemic, practical, moral, legal, etc. I will not address most of these concepts here; for further discussion, see \citet{Kment2021}.

\section{Counterintuitive Examples of the Received View of Physical Possibility}\label{section_threeexamples1}

The theory-relative formulation of the received view of physical possibility often leads to counterintuitive conclusions. Consider the following three examples:
\begin{itemize}
	\item[1.] Uncaused, indeterministic events are physically possible according to Classical Mechanics.
	\item[2.] G\"odel's time travel is physically possible according to General Relativity.
	\item[3.] Radical freedom of acting otherwise is physically possible according to Quantum Mechanics.
\end{itemize}

The first example might be surprising, as physicist folklore commonly regards Classical Mechanics as a paragon of determinism. However, both the mechanics of point masses satisfying surface constraints \citep{Norton2003a} and Newtonian gravitational theory of point masses \citep{Xia1992} provide examples where initial conditions do not uniquely determine a solution, thereby violating determinism. In Norton's example, Newton's equation for a point-like ball placed on the top of a specially shaped Dome has more than one solution: the ball can start rolling at an arbitrary time that is not determined by the shape of the Dome or the initial position and velocity of the ball. Norton uses this example of initial value indeterminism to argue further that our usual concept of causality is violated by Classical Mechanics, as motion may apparently start spontaneously without any cause.

The second example asserts that G\"odel's time travel is physically possible according to General Relativity. The law of General Relativity, Einstein's equation, is consistent with possible worlds in which some observers time travel (return to their earlier spacetime points after some time has elapsed). The first solution to Einstein's equation containing closed timelike curves, which allow for such time travel, was found by \citet{Godel1949}.  G\"odel's solution describes a static, non-expanding universe filled with a uniformly rotating homogeneous perfect fluid. Our actual universe is clearly not described correctly by G\"odel's solution, nevertheless \citet{Godel1949} famously argued that the existence of his solution implies that time itself is not real.

The third example requires a more detailed explanation. According to indeterministic interpretations of Quantum Mechanics, there are situations where multiple possibilities with non-negligible probabilities can occur. For example, a particle has a $\frac{1}{2}$ chance of decaying and a $\frac{1}{2}$ chance of not decaying during its half-life. If we think that quantum events with non-negligible probabilities can influence our actions in certain situations, then the essential condition for libertarian free will, the possibility of acting otherwise, is met in these situations. The fact that Quantum Mechanics, in this sense, is compatible with the possibility of acting otherwise is widely known.

However, it is not widely appreciated that the situation in Quantum Mechanics is even more radical. Suppose I glance at my partner (observe the positions of the particles making up their body at a given moment) and see that they are peacefully sleeping next to me in bed. The observed quantum state and the relevant physical properties of the room, through Schr\"odinger's equation, jointly determine the new quantum state two seconds later. This new quantum state, for instance and the sake of argument, could imply that the probability of finding my partner still peacefully sleeping upon a subsequent glance is more decimal places closer to 1 than the number of atoms in the universe. However, a consequence of time evolution according to Quantum Mechanics is that the probability of any single observational result (satisfying certain constraints) will not be precisely zero upon a subsequent glance. Therefore, there is a fantastically small probability that two seconds later my partner will laugh uproariously, jump on the chair, levitate above the bed, and so on. This means that all these ``actions" are consistent with both the law of Quantum Mechanics and the initial state. Thus, according to Quantum Mechanics, radical freedom of acting otherwise, understood in this sense, is physically possible, even though in the discussed particular situation the probability of any observed action other than continuing to sleep is so small that such events would never occur throughout the entire history of the universe. In other words, even though this situation strongly inductively compels that only one action is possible, multiple actions are physically possible according to Quantum Mechanics.

A common response to these and similar counterintuitive examples is that they rely on questionable idealizations \citep[see, e.g.,][]{Malament2008}. However, this response is problematic for two reasons. First, the relevant idealizations are routinely and successfully employed by the respective theories in other circumstances \citep{Norton2008}. Second, the response misunderstands the dialectical situation. The question of the theory-relative formulation of physical possibility is what possibilities arise {\em if} the given theory were true. When we question the problematic nature of the idealizations, we essentially deny this presupposition, i.e., we claim that the given theory is not true, only approximately true. However, denying the presupposition does not challenge the truth of the conditional statement.\footnote{While it is likely that no physicist today believes that Classical Mechanics is exactly true, the problem of initial value indeterminism appears in most modern physical theories; see \citet{Earman1986, Earman2009}.}

Of course, we can be dissatisfied with the dialectical situation itself, but it seems difficult to provide a better analysis of physical possibility of the received view than theory-relative physical possibility. Currently, there is no universally accepted, comprehensive theory of physics capable of explaining the phenomena that our best physical theories explain separately (String Theory is not considered a universally accepted physical theory, only a promising research program). Strictly speaking, General Relativity and the Standard Model of particle physics contradict each other, and thus the set of possibilities consistent with both theories is empty. Since a concept of physical possibility that implies no physically possible worlds exist is of little use, it seems we cannot do better than compare the results of physical possibility relative to different theories.

\section{Logical and Empirical Possibility}\label{section_logicalempirical}

\citet{Norton2022} presents a robust critique of the philosophical literature on modality. He contends that there are only two defensible concepts of possibility and necessity: the logical and the empirical. According to Norton, all other instances of modality either reduce to these two concepts or are not cogent and cannot be responsibly relied upon.

Norton understands logical possibility differently from its standard usage in the literature, which may lead to misunderstandings. The standard concept of logical possibility appears in modal logic, where a proposition is logically possible if it is true in at least one possible world. In non-modal logic, the concept of logical possibility is not clearly defined, but in this context, a proposition is usually called logically possible if its negation is not a tautology, or if its negation is not a logical truth, or if, under the axioms and inference rules of the given logic, the proposition does not lead to a contradiction.

Norton understands logical possibility in a broader way, relativizing it to a set of further propositions $L$. \citet[132]{Norton2022} defines a proposition $S$ to be {\em logically possible relative to a set of propositions} $L$ if $S$ and $L$ are logically consistent (in this paper, I will use the notation $\Diamond^l_L S$ to denote $L,S \nvdash \bot$). The third above-mentioned logical possibility concept of non-modal logic is thus a special case of Norton's logical possibility concept, namely when the set $L$ is empty. Norton defines a proposition $S$ {\em logically necessary relative to a set of propositions} $L$ if $S$ deductively follows from this set $L$ (in this paper, I will use the notation $\Box^l_L S$ to denote $L \vdash S$). Since $L \vdash \neg S$ if and only if $L, S \vdash \bot$, it is clear that the usual duality holds between Norton's logical possibility and logical necessity: $\Box^l_L \neg S = \neg \Diamond^l_L S$.

I am not aware of other philosophers using the concepts of logical possibility and necessity in the same way as Norton. The typical distinguishing feature of logical possibility is that it {\em only} requires consistency with the axioms of a given logic, without requiring consistency with any additional, non-logical axioms or propositions. However, for Norton, the distinction between logical and non-logical axioms is secondary (after all, one can obtain a new logic by incorporating the set $L$ of propositions into the logical axioms). For Norton, the key question is whether we are working with a concept of possibility that only requires steering clear of logical contradiction, or with a concept that defines the scope of possibilities in a manner different from the mere requirement of avoiding logical contradiction. Norton's logical possibility captures the former, while his empirical possibility exemplifies the latter type.

Norton's concepts of empirical possibility and necessity are captured by the following concise definition (ibid. 134): a contingent proposition $S$ is {\em empirically possible} if the actual body of evidence $A$ positively inductively supports $S$, and $S$ is {\em empirically necessary} if the actual body of evidence $A$ inductively compels $S$. In this paper, I will use the notations $\Diamond^e_A S$ and $\Box^e_A S$ for Norton's empirical possibility and necessity.

The most important element in defining empirical possibility and necessity is the concept of inductive support. Here, Norton relies on the material theory of induction, which he has developed in several books and articles over recent years \citep[see][]{Norton2003b, Norton2021, Norton2024}. Although a detailed presentation of the material theory of induction is beyond the scope of this paper, the basic idea is well illustrated by the following example \citep[649]{Norton2003b}. Consider two cases of enumerative induction:
\begin{itemize}
	\item[(P1)] The melting point of several samples of bismuth is 271$^o$C.
	\item[(C1)] Therefore, the melting point of all bismuth is 271$^o$C.
	\item[(P2)] The melting point of several samples of wax is 91$^o$C.
	\item[(C2)] Therefore, the melting point of all wax is 91$^o$C.
\end{itemize}
Although the formal structure of the two enumerative inductions is the same, (P1) $\rightarrow$ (C1) is a strong inductive inference, while (P2) $\rightarrow$ (C2) is a weak inductive inference. What might explain this difference in strength? According to Norton, the difference in their strengths is due to the fact that the first induction is licensed by a material fact obtained as the result of another, earlier successful inductive inference (namely, the fact that samples of chemical elements generally have fixed melting points), while no such material fact licenses the second induction (as amorphous substances like wax generally do not have fixed melting points).

Norton argues that the situation illustrated by the examples of bismuth and wax generalizes: the strength of an induction is always based on some material fact related to the actual content of the induction, rather than on formal properties of the inductive facts or inferences. Although Norton acknowledges that inductive inferences may have numerical degrees of strength, he criticizes the Bayesian approach for attempting to universally reproduce this gradation and to provide a formal theory of inductive inferences. Norton argues that no such general formal theory of induction exists.

A natural question arises as to how we arrive at the material facts necessary to support an induction (e.g., the material fact that samples of chemical elements generally have a fixed melting point). According to Norton, these material facts are often themselves results of previous inductive inferences, supported by even earlier established material facts. This leads to a regression but Norton argues that the regression is not paradoxical; by tracing back inductions we are essentially uncovering the history of science.

According to Norton's arguments and intentions, a material induction has the following characteristics.
\begin{itemize}
	\item[1.] {\em Relational}: The inductive relationship between premises and conclusions can only be understood in relation to further material facts (generally, further actual evidence).
	\item[2.] {\em Gradational}: The inductive relationship between premises and conclusions can have varying degrees of strength.
	\item[3.] {\em Non-epistemic}: Induction is an objective relationship between premises, conclusions, and actual evidence, independent of agents' belief and thoughts.
	\item[4.] {\em Language-independent}: Although Norton describes induction as a relationship between propositions, he emphasizes that induction does not inherently depend on the language in which the relevant contingent facts are formulated.
\end{itemize}
Since inductive support is a key element of empirical possibility and necessity, these concepts inherit the characteristics of material induction mentioned above. \citet{Norton2022} thus argues that empirical possibility and necessity are themselves relational, gradational, non-epistemic, and language-independent concepts. For example, empirical possibility should not be confused with epistemic possibility because it is not defined through the ignorance of agents. The difference between empirical and epistemic possibility is evident, for instance, when the actual body of evidence $A$ is assumed to be empty (or very sparse); in such a case, no proposition $S$ is empirically possible because $A$ cannot provide positive inductive support for any proposition. On the contrary, with an empty body of evidence $A$, every proposition $S$ is epistemically possible for an agent, since the agent does not know about any proposition that it does not hold (ibid. 146---147).

The requirement for positive inductive support for empirical possibility and inductive compulsion for empirical necessity are posited independently. A direct consequence of this is that duality is typically violated between empirical possibility and necessity: in general, $\Box^e_A S \neq \neg \Diamond^e_A \neg S$! Norton argues extensively that duality can only be expected to hold under special circumstances. The violation of duality also implies that possible world semantics is, in general, not applicable to empirical possibility and necessity. Therefore, according to Norton, possible world semantics must generally be abandoned.

Another important characteristic of empirical possibility and necessity is the reliance on actual evidence. To emphasize the empirical roots of these concepts, Norton stipulates that the body of evidence $A$ must be actual, which means that it includes only evidence derived from actual experience. Norton uses this {\em condition of actuality of evidence} to exclude a concept of empirical possibility based on imagined experiences in hypothetical worlds: our actual experiences derive from our actual world, not from the imagination of metaphysicians. (In Section \ref{section_condposs}, we will return to Norton's concepts of hypothetical and counterfactual possibility, which relax the condition of actuality of evidence.)

Norton's phrasing at several points suggests that the body of evidence $A$ in the definitions of empirical possibility and necessity includes {\em all} actual evidence; however, he does not explicitly state this assumption in the article. It is apparent that the body of evidence changes over time and was different in Newton's time, for example, than it is today. Nevertheless, Norton's intention is clear that the body of evidence $A$ must be sufficiently rich to allow the application of the material theory of induction. Since, according to the material theory of induction an inductive generalization must be supported by material facts which are also consequences of previous inductions and earlier material facts, it is clear that if $A$ includes any inductive generalizations, it must also include the material facts supporting these generalizations and earlier inductions supporting those material facts, and so on. Hence, few plausible candidates for $A$'s role exist other than the total actual body of evidence of a given era and domain of inquiry.

Thus, Norton's concept of empirical possibility is fixed by its position on two, independent axes. On the first axis we have two options: logical vs. inductive, and the question is about the type of relationship a proposition $S$ has with another set of propositions $L$: if $S$ is logically consistent with $L$ then $S$ is logically possible relative to $L$, while if $S$ is positively inductively supported by $L$ then $S$ is inductively possible relative to $L$. The second axis determines the relationship of the set of propositions $L$ to the total body of actual evidence $A$; $L$ may coincide with $A$, may contain only certain parts of $A$, or may be disjoint from $A$ altogether. Norton says that $S$ is empirically possible if, according to the first axis, $S$ is positively inductively supported by $L$ and if, on the second axis, $L$ coincides with the total body of evidence $A$. As we will see later, Norton's concepts of hypothetical and counterfactual possibility are also inductive (first axis), but they relax the condition that $L$ coincides with the total body of evidence $A$: $L$ may only contain certain parts of $A$ (second axis). However, $L$ needs to include a part of $A$ substantial enough to allow for the use of the material theory of induction. (Naturally, we may also ask what the relationship of $L$ with total body of actual evidence $A$ is for the case of logical possibility: $L$ may or may not contain certain parts of $A$. But the answer to this second question (second axis) is independent of whether $S$ is logically consistent with $L$, that is, whether $S$ is logically possible relative to $L$  (first axis), according to Norton's definition of logical possibility.)

Armed with the concepts of logical and empirical possibility and necessity, \citet{Norton2022} aims to show that all other modal concepts can only be retained if they can be subsumed in one of the logical or empirical variants. Norton's primary targets of criticism are epistemic and metaphysical possibility and necessity. While his arguments are exciting and entertaining, I do not reconstruct them here; my main focus is on the relationship between Norton's concepts of logical and empirical possibility and the received view of physical possibility.

\section{Physical Possibility: Empirical or Logical?}\label{section_emporlogical}

\noindent Norton makes the following claim about the concept of physical possibility:
\begin{quote}
Physical possibility refers to mundane possibilities learned through the routine operation of science. They are empirical possibilities. \citep[145]{Norton2022}
\end{quote}
Thus, Norton claims that physical possibilities are empirical possibilities. In this section, I examine whether this claim is tenable if by physical possibility we understand the received view introduced in Section \ref{section_physicalpossibility}. To avoid unnecessary repetition, let me now stress clearly:  by the term physical possibility I mean physical possibility of the received view, unless explicitly stated otherwise.

Identifying physical possibilities as empirical possibilities raises several immediate issues. As I mentioned earlier, duality is generally violated between Norton's concepts of empirical possibility and necessity: $\Box^e_A S \neq \neg \Diamond^e_A \neg S$. However, duality holds between the concepts of physical possibility and necessity: $\Box^p_T S = \neg \Diamond^p_T \neg S$! Moreover, while possible worlds semantics is generally not applicable to empirical possibility and necessity, the concepts of physical possibility and necessity are defined using possible worlds semantics!

One might argue that duality could hold for the special case of empirical possibility that constitutes physical possibility; such an assumption would not contradict Norton's analysis. However, it is unclear -- and Norton's analysis does not address this question -- why physical possibility should be a special case of empirical possibility in this sense.

Let us proceed. According to Norton's claim, for a proposition $S$ if  $\Diamond^p_T S$ then $\Diamond^e_A S$, where $A$ is the actual body of evidence. What, then, could be the theory $T$ in this formula according to which $S$ is physically possible? Norton provides no indication which physical theory's laws are relevant to obtain empirical possibilities. The apparent options, however, do not seem very appealing.

We might assume that the sought-after theory $T$ could be any inductively supported physical theory. Depending on the required degree of support, we may obtain a varied picture of what counts as empirically possible. Returning to the three examples from Section \ref{section_threeexamples1}: if Classical Mechanics counts as sufficiently supported, uncaused, indeterministic events would be empirically possible according to it. Similarly, if General Relativity counts as sufficiently supported, G\"odel's time travel would be empirically possible according to it. Finally, if Quantum Mechanics counts as sufficiently supported, radical freedom of acting otherwise would be empirically possible according to it. Therefore, empirical possibility would support the same counterintuitive conclusions as theory-relative physical possibility.

The main problem with allowing any inductively supported current physical theory to play the role of $T$ is the fact mentioned earlier that even our best-supported current physical theories contradict each other: there are situations (for example, how particles exhibiting quantum behavior behave in strong gravitational fields) where General Relativity and the Standard Model make contradictory propositions (Norton himself agrees with this, see ibid. 148, footnote 21). Thus, with some choices of $T$, certain situations would be considered empirically possible, while other choices would deem the same situations not empirically possible, thereby rendering the concept of empirical possibility contradictory.

The same contradiction discussed in the previous paragraph would arise if we took the union of currently supported physical theories as the sought-after theory $T$.

If we assume that the sought-after theory $T$ is a future accepted physical theory that successfully reconciles the currently supported physical theories without contradiction, the problem is different: the content of this future theory is currently unknown, and there is no guarantee that the propositions which the future theory would hold as possible can even be formulated based on our current body of evidence $A$. For example, suppose that this future unified theory of physics posits that the fundamental entities of our world are 93-dimensional springs and asserts claims such as, ``93-dimensional springs are arranged in this and that way." The current body of evidence $A$ says nothing about the existence of 93-dimensional springs, so it is clear that $A$ cannot provide positive inductive support for such propositions, making them not empirically possible (thus, $\Diamond^p_T S$ but not $\Diamond^e_A S$). According to the well-known pessimistic meta-induction, we should expect that our future, currently unknown physics indeed posits entities not even hinted at by our current body of evidence \citep[cf.][]{Shech2019}.

The same issue discussed in the previous paragraph would arise if we took the theory whose laws match those of our actual world as the sought-after theory $T$. The physical laws of the actual world are also currently unknown to us.\footnote{The above line of reasoning is similar to the so-called Hempel's dilemma. \citet{Gyenis2022} argues that the dilemma can be resolved.}

Therefore, we have not succeeded in finding a physical theory that, when chosen, would convincingly support the claim that physical possibilities according to this theory are empirical possibilities. I would like to illustrate why finding such a physical theory is unlikely with a simple example. Every physical theory that was at some point considered fundamental since the 17$^{th}$ century formulated its dynamic physical laws in the form of differential equations. These differential equations have solutions naturally interpreted as worlds devoid of any matter, ``empty". Rephrased, according to the dynamic laws of these theories, the proposition ``the world is empty" expresses a physically possible states of affairs. However, the empty world cannot be empirically possible, as the total body of evidence contains many actual experiences incompatible with the emptiness of the world. Thus, the proposition ``the world is empty" is physically possible according to the dynamic laws of every fundamental physical theory, yet it is not empirically possible. Therefore, not all physical possibilities can be empirical possibilities.\footnote{Some philosophers believe that not only dynamic laws can be fundamental physical laws, but also certain generalizations about initial or boundary conditions, such as the so-called Past Hypothesis, which posits that the universe began in a state of low (but non-zero) entropy. While positing the Past Hypothesis to be a physical law excludes the physical possibility of an empty world, the Past Hypothesis is evidently still compatible with many physically possible worlds that are not consistent with the totality of our actual body of evidence.} (Section \ref{section_condposs} will return to the example in more detail.)

The physical possibility of an empty world highlights a property of the received view of physical possibility, which I call the {\em problem of separating evidence}. The received view of physical possibility captures the idea of possibilities permitted by the laws of nature: the core idea is to separate what is considered essential in physically possible worlds (namely, the laws) from what is considered contingent (such as the distribution of matter, generally, the initial and boundary conditions). For physical possibility according to some theory to coincide with empirical possibility based on a body of evidence, the body of evidence underlying the empirical possibility would need to include only the evidence necessary to inductively support the laws and exclude evidence pertaining to contingent facts about our actual world. However, it seems clear that we cannot separate the evidence that only supports the laws from the evidence that supports contingent facts. Inductive support for any law depends on conducting experiments or making observations in specific situations. All these specific situations contain information not only about the laws of our actual world but also about its contingent structure. Consequently, there must be numerous possibilities permitted by the laws that are not consistent with the body of evidence positively inductively supporting the laws, making these physical possibilities not empirically possible.\footnote{Another issue relates to Norton's material theory of induction. Let $T$ now be the maximal physical theory whose physical possibilities imply empirical possibilities. If this maximal $T$ is also a physical theory where the implication is mutual (if for every proposition $S$, $\Diamond^p_T S$ if and only if $\Diamond^e_A S$), then $T$ would provide a formal theory of induction, as physical theories are formalized and consistency with $T$'s physical laws is a formal requirement. This would contradict Norton's main claim that no formal theory of induction exists. If, on the other hand, there is an empirical possibility that is not a physical possibility according to this maximal $T$ theory (if there is an $S$ such that $\Diamond^e_A S$ but not $\Diamond^p_T S$), then we have two options. Either we must conclude that such an empirical possibility is merely epistemic, which then contradicts Norton's claim that empirical possibility is not epistemic possibility. Alternatively, we must conclude that Norton's account implies a form of anti-physicalism, since it posits the existence of a non-epistemic possibility that is not a physical possibility even according to the maximal physical theory.}

In summary, it is unclear how Norton's claim that physical possibilities are empirical possibilities could be true, unless Norton makes a claim about his own (undefined) notion of physical possibility, as opposed to the received view of physical possibility. Although Norton's claim is rather terse, he addresses the relationship between hypothetical and counterfactual possibility and nomic possibility in more detail. I will return to hypothetical and counterfactual possibilities in Section \ref{section_condposs}.

If physical possibilities of the received view are not empirical possibilities, then where do they fit into Norton's map of modalities?

At this point, it is worth noting that if the set of propositions $L$ coincides with the laws of $T$, then according to Norton, the propositions {\em logically} possible relative to $L$ correspond to the {\em physically} possible propositions in the non-modal sense, as discussed in Section \ref{section_physicalpossibility}, that is (non-modally) $\Diamond^l_L S \leftrightarrow \Diamond^p_T S$! It is thus clear that the non-modal formulation of physical possibility given by Chisholm is a special case of Norton's non-modal logical possibility concept.

The parallel between Norton's and Chisholm's formulations in the non-modal case can also be straightforwardly extended to the modal case. As we have seen, Chisholm's non-modal formulation of physical possibility is a special case of the modal formulation (p). Norton's non-modal logical possibility can also be naturally generalized to a modal logical possibility: a proposition $S$ is logically possible relative to a set of propositions $L$ if $S$ is true in at least one possible world consistent with the propositions $L$. With this generalization, the formulation (p) of physical possibility becomes a special case of Norton's (modalized) logical possibility: if the set of propositions $L$ coincides with the physical laws of theory $T$, then (modally) $\Diamond^l_L S \leftrightarrow \Diamond^p_T S$!

In summary, the received view of physical possibility is a special case of Norton's (modalized) logical possibility. Since, according to Norton, logical possibility differs from empirical possibility, we have a final argument as to why physical possibility of the received view cannot be empirical possibility: empirical possibility differs from logical possibility, and physical possibility of the received view is just a case of Norton's logical possibility.

\section{Empirical Possibility and Counterintuitive Examples of Physical Possibility}\label{section_threeexamples2}

\noindent To illustrate the difference between empirical and physical possibility, let us return to the examples of Classical Mechanics, General Relativity, and Quantum Mechanics. 
\begin{itemize}
\item[1] In the domain of classical physics: \\
(a) The proposition that a stone ball placed on the wheelchair ramp of the Cathedral of Learning begins to roll downwards is physically possible according to Classical Mechanics, as it is consistent with Newton's law.  \\
The same proposition is also empirically possible, as the totality of our current evidence provides positive inductive support for it. This is a typical situation in which we have ample evidence that Classical Mechanics provides accurate predictions, and the prediction of Classical Mechanics is that the stone ball will begin to roll downwards. \\
(b) The indeterminism illustrated by Norton's Dome or Xia's point masses is physically possible according to Classical Mechanics, since the solutions to Norton's and Xia's models are consistent with Newton's law. \\
However, the initial value indeterminism illustrated by Norton's Dome or Xia's point masses is not empirically possible. Both Norton's and Xia's conclusions about indeterminism rely on idealizations (in Norton's case: balls can be held on a surface with arbitrary force \citep[see][]{Malament2008}; in Xia's case: the particles have no extension, and there is no limiting speed) that cannot be relaxed without losing Classical Mechanics' conclusions about indeterminism. These idealizations are also such that current evidence (from well-established areas of physics, such as solid-state physics, quantum physics, and relativistic physics) compellingly indicates that they do not hold: there are no arbitrary surface forces, no real point masses, but a limiting speed exists. 
\item[2] In the domain of relativistic physics: \\
(a) The proposition that after my round-trip flight next week the difference between my biological age and that of my siblings will slightly increase is physically possible according to General Relativity, as it is consistent with Einstein's equation. \\
The same proposition is also empirically possible, as the totality of current evidence provides positive inductive support for it. This is a typical situation in which we have ample evidence that General Relativity provides accurate predictions (i.e. from the 1971 experiments of Hafele and Keating), and the prediction of General Relativity is that a slight increase in our biological ages will occur as a result of the trip. \\
(b) Time travel of the G\"odel's universe is physically possible according to General Relativity, as G\"odel's universe is a solution of Einstein's equation that contains closed timelike curves. \\
However, time travel of the G\"odel's universe is not empirically possible, as the total body of evidence compels that G\"odel's universe does not accurately represent our actual world.
\item[3] In the domain of quantum physics: \\
(a) The proposition that my partner, peacefully resting in bed, will still be asleep two seconds later is physically possible according to Quantum Mechanics. Schr\"odinger's equation is consistent with finding that after measurement a quantum state (that instantiated a sleeping person two seconds earlier) still instantiates a sleeping person. \\
The same proposition is also empirically possible, as supported by everyday experiences. \\
(b) The proposition that my partner, peacefully resting in bed, will be levitating above the bed two seconds later is physically possible according to Quantum Mechanics. Schr\"odinger's equation is consistent with my sleeping partner entering a quantum state in which there is an extraordinarily small probability that a subsequent measurement finds my partner levitating above the bed. \\
However, the same proposition is empirically necessarily false. According to the assumption, the mentioned probability is so fantastically small that the total body of evidence inductively compels such an outcome not to happen. \\
(According to Norton, the duality between empirical necessity and possibility fails, and thus it may be the case that a proposition is empirically necessarily false yet empirically possible. However, empirical possibility itself still requires {\em positive} inductive support. If the probability is sufficiently small, as in the example, and especially if it is zero, positive inductive support is not forthcoming, hence the proposition is also not empirically possible.)
\end{itemize}

The above examples reinforce the main conclusion of the previous section: physical possibilities often do not coincide with empirical possibilities. The differences appear to arise in three areas where the concept of physical possibility, at least from the perspective of a practicing physicist, leads to counterintuitive conclusions.

The example from Classical Mechanics illustrates that while the theory-relative formulation of physical possibility struggles with the problem that different accepted physical theories, speaking from a strict logical standpoint, can imply contradictory physical possibilities, Norton's empirical possibility can sometimes resolve such contradictions, even without an appeal to a future unified theory of physics. The total body of evidence already includes propositions regarding the domains of validity of physical theories. Thus, if the source of the contradiction is that some theories would be applied outside their domains of validity, then these applications would be halted by inductive inferences that are based on the total body of evidence. 

The example from General Relativity illustrates that, while the theory-relative formulation of physical possibility merely requires logical consistency with the laws, Norton's empirical possibility depends not only on the laws but also on other non-nomic facts originating from actual experiences. From an empiricist perspective, it is not entirely clear on what basis we can attribute a special, higher status to a strongly inductively supported law over a strongly inductively supported non-nomic fact (see the quote from Norton in the next section). If we do not attribute special status to laws, then the concept of empirical possibility (or at least Norton's logical possibility relative to some non-nomic facts in addition to the laws) provides a more adequate framework than the received view of physical possibility.

Finally, the example from Quantum Mechanics illustrates that the concept of empirical possibility can capture the phenomenon that in case a theory implies that the probability of an alternative is negligibly small or zero, physicists often phrase this as saying that the alternative is not possible according to the theory. Although this manner of speaking can be criticized as a conflation of small or zero probability with impossibility, the phrasing aligns with the criterion that empirical possibility requires positive inductive support.

\section{Types and Degrees of Conditional Inductive Possibility}\label{section_condposs}

\noindent In the previous section, I concluded that physical possibility of the received view cannot coincide with Norton's empirical possibility. In the next two sections I argue that Norton's framework allows for a more nuanced handling of counterintuitive examples of the received view.

In Section 11 of his article titled ``Nomic Possibility," Norton further analyzes the possibility concept germane to the sciences. His analysis is concise, allowing us to quote a substantial portion of it.
\begin{quote}
First, I do not accord scientific laws any special status qualitatively in comparison with other contingent facts. They are not some higher order of truth beyond ordinary contingent truths. They are merely contingent truths of very broad scope. None that we know has universal application. Thus they can afford us no notion of possibility and necessity that is distinct from that afforded by ordinary contingent truths. They merely do so with greater scope. \\
$[$...$]$ \\
The modal aspect of scientific theories is, when examined more closely, captured fully with inductive notions. $[$...$]$ In my rendition, a scientific theory is simply a large collection of possibility and necessity claims, often encoded elegantly in quite compact statements. In their domain of application, they assert propositions of the form ``If this happens, then that may possibly ensure, or that other may necessarily ensue, but that other again necessarily cannot ensue." Propositions of this form are inductive statements. They can be restated as ``If this happens, then there is some evidence for that; and compelling evidence for that other, etc." That is, they are large collections of propositions that lie within the compass of empirical or logical possibility. Since most of the ``this's" will not describe our actual evidence, these propositions almost all assert hypothetical or counterfactual possibilities (as defined in Sect. 4 above). There is nothing primitively modal about them. \citep[148]{Norton2022}
\end{quote}
The hypothetical and counterfactual possibilities mentioned in the quote derive from the concept of empirical possibility by relaxing the condition of actuality of evidence:
\begin{quote}
There is a place for hypothetical and counterfactual possibilities in the account if we relax this evidential actuality condition. A {\em counterfactual possibility} is defined as one that derives inductive support from a body of evidence in contradiction with the evidence provided by experience. A {\em hypothetical possibility} is defined as one that derives inductive support from a body of evidence not fully recovered from experience but logically compatible with it. (ibid. 137, emphasis in the original.)
\end{quote}

To facilitate later analysis, I now reconstruct Norton's concepts of counterfactual and hypothetical possibilities as follows. In harmony with the notation used so far, let the capital Latin letter $A$ denote the total actual body of evidence derived from experience. According to Norton's definition, a proposition $S$ is hypothetically possible if it is inductively supported by {\em a} body of evidence that is not {\em fully} recovered from experience. This means that we can divide a body of evidence inductively supporting $S$ into two parts: one part that follows compellingly from the total body of actual evidence $A$ -- denoted $A^{-}$ --, and another part that does not follow compellingly from $A$ -- denoted $H$. Thus, we assume that $S$ is positively inductively supported by $A^{-} \cup H$, and that $\Box^e_A A^{-}$, and $\neg \Box^e_A H$. With these assumption, $S$ is {\em hypothetically possible} according to Norton if $H$ is logically compatible with $A$ (if $\Diamond^l_A H$), and $S$ is {\em counterfactually possible} if $H$ logically contradicts $A$ (if $\neg \Diamond^l_A H$).

Based on this reconstruction, it is easy to see that Norton's empirical, hypothetical, and counterfactual possibilities are special cases of what I call conditional inductive possibility. A proposition $S$ is {\em conditionally inductively possible with respect to $A$, $A^{-}$, and $H$} (in notation: $\Diamond^c_{A, A^{-},H} S$) if $S$ is positively inductively supported by $A^{-} \cup H$, where $A^{-}$ is a non-empty body of evidence compelled by the total actual body of evidence $A$ ($\Box^e_A A^{-}$), and $H$ is either empty or the total actual body of evidence does not compel $H$ ($\neg \Box^e_A H$). With this definition a conditionally inductively possible proposition $S$ is empirically possible if $H$ is empty and $A^{-}$ coincides with $A$ ($A^{-} = A$), partially empirically possible if $H$ is empty and $A^{-}$ does not coincide with $A$ ($A^{-} \neq A$), hypothetically possible if $H$ is non-empty and is logically possible according to $A$ ($\Diamond^l_A H$), and counterfactually possible if $H$ is non-empty and is not logically possible according to $A$ ($\neg \Diamond^l_A H$).

Two remarks about the relationship between logical and inductive modalities are in order. Norton emphasizes the following properties:
\begin{quote}
Empirical necessity includes logical necessity as a limiting case. [...] However logical compatibility without inductive support is not included as a limiting case of empirical possibility. \citep[135]{Norton2022}
\end{quote}
First, note that any proposition $S$ trivially deductively follows from $A^{-} \cup S$; since, according to Norton, logical necessity is a limiting case of empirical necessity, $A^{-} \cup S$ also then inductively compels $S$, and thus $A^{-} \cup S$ also positively inductively supports $S$.  Therefore, if we condition on $S$, then $S$ is counterfactually possible (for all $S$: $\Diamond^c_{A, A^{-},S} S$). This is clearly a trivial and uninteresting case of counterfactual possibility. The same situation holds when, instead of $S$, we condition on a proposition $H$ such that $S$ deductively follows from $A^{-} \cup H$: in this case $S$ is again counterfactually possible, but again only in the trivial sense that if I essentially condition upon $S$ (since I condition upon premises from which $S$ deductively follows) then, obviously, $S$ becomes counterfactually possible. To avoid trivializing the discussion, henceforth I shall assume that we understand conditional inductive possibility in a non-trivial way (that is, we assume that $\neg \Box^l_{A^{-} \cup H} S$).

Second, noting that conditional inductive possibility is a generalization of empirical possibility, the case when a proposition is merely logically compatible with a body of evidence without being positively inductively supported by said body of evidence (that is, when a proposition $S$ is merely logically compatible with $A^{-} \cup H$ without $S$ being positively inductively supported by $A^{-} \cup H$) can {\em not} taken to be an instance of empirical, hypothetical, or counterfactual possibility, in general: can not taken to be an instance of conditional inductive possibility. Clearly, such cases have no connection with the material theory of induction and are simply instances of Norton's {\em logical} possibility.

Let us return now to the relationship of nomic possibility and conditional inductive possibility. In the second part of the first quote, Norton states that nomic possibilities are hypothetical or counterfactual possibilities. This claim cannot hold for the received view of nomic possibility. Nomic possibility of the received view, in complete analogy with physical possibility, is a special case of Norton's logical possibility. However, the intended and interesting cases of conditional inductive possibility are those in which the material theory of induction plays an actual role in determining what is possible. But then nomic possibilities cannot coincide with conditional inductive possibilities. The argument parallels that of Section \ref{section_emporlogical}: duality of possibility and necessity is violated for conditional inductive possibility the same way as it is violated for empirical possibility and necessity (nomic possibility and necessity, however, do satisfy duality), and the problem of separation of evidence plagues nomic possibility the same way as it plagues physical possibility. In general, it would be a surprising consequence of Norton's account if the correct reconstruction of ``almost all" possibility claims of the sciences invoked logical possibility instead of inductive possibility. If anywhere, it is in the empirical sciences that we should expect the reconstruction of possibility talk to involve conditional inductive possibility, in which the material theory of induction plays an actual role. What other area would benefit from conditional inductive possibility if not how science works?

To reiterate my earlier illustration, an empty world that is nomically possible according to an accepted scientific theory is still not a conditional inductive possibility. Consider General Relativity. Minkowski spacetime is a solution to Einstein's equation. Since Minkowski spacetime can represent a completely empty universe, the statement $S$ which asserts that ``the universe is completely empty," expresses a state of affairs which is physically (nomically) possible according to General Relativity. In other words, if $E$ includes Einstein's equation (and all the mathematical axioms necessary for its formulation, but nothing else), then, since $S$ is consistent with $E$, in Norton's sense $S$ expresses a logically possible situation relative to $E$.

Now let us ask: is the set of mathematical statements $E$, which contains only Einstein's equation, sufficient to positively inductively support the claim $S$ that the universe is completely empty, according to the material theory of induction? The answer must clearly be negative: a mere mathematical equation constraining dynamical evolution cannot, on its own, provide positive inductive support for a completely independent claim regarding the actual distribution of matter. $E$, by itself, is not rich enough to positively inductively support such claims, not least because it lacks material facts upon which an inductive inference could rely.

For the material theory of induction to be applicable, it is necessary to complement $E$ with additional facts compelled by the totality of our actual experiences, such that the enriched set $A^{-}$ (containing $E$) becomes sufficiently rich to positively inductively support claims about the actual distribution of matter. At a minimum, $A^{-}$ must include empirical, material facts that enable the symbols in $E$ to acquire physical meaning and be connected to claims about the distribution of matter. These are essentially the empirical facts that underpin our understanding of General Relativity as a physical theory, as learned in universities and through laboratory experiments. However, since an $A^{-}$, which includes such facts, logically entails the existence of universities, laboratories, and humans, it also logically follows that $S$, the claim that the universe is completely empty, is false. Consequently, such an $A^{-}$ cannot provide positive inductive support for $S$, and thus $S$ cannot express an empirical, hypothetical, or counterfactual possibility in Norton's sense.\footnote{My question concerned whether the states of affairs described by statement $S$, asserting that the universe is completely empty, is physically and/or conditionally inductively possible. Physical possibility followed from the conventional understanding of Minkowski spacetime as a spacetime that {\em can} represent a completely empty universe. This is independent of the question of whether Minkowski spacetime, as a mathematical structure, can, with sufficiently loose approximations and for certain purposes, {\em also} be used to represent part (or even the entirety) of our actual universe which does contain matter. I have not argued that we cannot have conditional inductive support for the statement that Minkowski spacetime could (approximately) represent part (or even the entirety) of our actual universe. Rather, I argued that the claim that our actual universe is completely empty cannot receive conditional inductive support.}

Naturally, $E$ combined with additional assumptions regarding the distribution of matter somewhere could potentially suffice for deductively drawing conclusions about the distribution of matter elsewhere. However, such inferences would have nothing to do with induction. For instance, if we assume, in addition to $E$, that spacetime is flat and empty on a Cauchy surface, then from this initial condition and $E$ it would deductively follow that the entire spacetime is Minkowski, making $S$ logically necessary. However, this is nothing more than the trivial and uninteresting case of conditional inductive possibility discussed earlier: of course, if I counterfactually assume that something holds, then it will follow that it is possible. Moreover, our original question was not whether Einstein's equation together with certain initial conditions render a completely empty universe conditionally inductively possible. Rather, the question was whether a completely empty universe is conditionally inductively possible according to General Relativity. While Einstein's equation is inductively supported by the totality of actual experiences, General Relativity does not include the (to our knowledge, false) explicit assumption that in our actual universe spacetime is flat and empty on a Cauchy surface.

Let us now return to the first part of Norton's quote. According to Norton, laws are merely propositions capable of elegantly encoding if-then propositions linking specific, contingent premises and conclusions -- initial conditions and outcomes. These specific if-then propositions express inductive relationships. The following hypothetical if-then proposition could be an example of the form Norton has in mind: let $H$ describe a hypothetical ball at the top of a slope, $S$ state the outcome that the ball starts rolling down, $A$ be the actual body of evidence, $A^{-}$ be the actual evidence pertaining to the behavior of balls; $\Diamond^c_{A, A^{-},H}$ expresses the conditional inductive possibility given $A$, $A^{-}$, and $H$. Thus $\Diamond^c_{A, A^{-},H} S$ states that if the ball is in the initial position described by $H$, then $H$ together with the $A^{-}$ body of evidence pertaining to the behavior of balls gives positive inductive support to the outcome described by $S$. According to Norton, we tend to interpret inductive relationships expressed by if-then propositions in modal terms, along the lines of empirical possibility and necessity. The modal reading of this example would be: if a ball is in the initial position described by $H$, then it is (conditionally inductively) possible that it rolls in the direction $S$.

The quotations from Norton suggest a reductive view of the inductive support of scientific laws and the modalities they provide. The reductive view suggests that inductive support primarily occurs at the level of particular if-then propositions linking initial conditions and outcomes. Since laws merely elegantly encode these if-then propositions in compact statements, the inductive support of laws (and hence the modalities they provide) is secondary, derived from the inductive support of particular if-then propositions.

The reductive view suggested by Norton allows us to assign different modalities to different if-then propositions that are nevertheless consistent with the same laws. The received view of physical possibility and necessity cannot provide such differentiation: if we assume that a given theory's physical laws are true, the only relevant question for physical possibility is whether a specific if-then proposition is consistent with the laws. If it is consistent, it is physically possible according to the given theory; if it is not, it is not. Beyond this binary determination, the received view of physical possibility does not provide a means to say something more detailed about why we might trust some if-then propositions more than others.

Within Norton's framework, there are two ways to assign different degrees of strength to different conditional propositions that are nevertheless consistent with the same laws. The first way is straightforward: inductive support is gradational and hence conditional possibility already comes in degrees. Thus, some if-then propositions are more or less conditionally inductively possible than others. 

 The second way is suggested by the following observation: for hypothetical and counterfactual possibilities, Norton only imposes a constraint on whether the hypothesis $H$, which is not compelled by the actual body of evidence $A$, is logically consistent with $A$. If $H$ is consistent with $A$, the possibility is hypothetical, if it is not consistent, the possibility is counterfactual. However, Norton does not impose a constraint on the degree to which hypothesis $H$ is empirically supported by $A$ (beyond the fact that it is not compelled by $A$, for such evidences by definition belong to $A^{-}$). Introducing such a constraint leads to a natural extension of Norton's account. I will follow this line of thought here.

\begin{table}
\centering
\caption{{\bf Types and Degrees of Conditional Inductive Possibility}}\label{table1}
\begin{tabular}{|l|}
\hline
$\Diamond^c_{A, A^{-},H} S$ if: $S$ is positively inductively supported by $A^{-} \cup H$, \\ where $A$ is the total actual body of evidence, $\Box^e_A A^{-}$, and $H$ is either empty or $\neg \Box^e_A H$.  \\ 
$\Diamond^c_{A, A^{-},H} S$ is {\em trivial} if $S$ deductively follows from $A^{-} \cup H$ (if $\Box^l_{A^{-} \cup H} S$). \\
\hline
Degrees of conditional inductive possibility: \\
Firstly, due to induction being gradational, $\Diamond^c_{A, A^{-},H} S$ itself can vary in strength. \\
Secondly, even with $\Diamond^c_{A, A^{-},H} S$ having the same strength, the hypothesis $H$ \\
itself can be supported to varying degrees. \\
Let $\Diamond^c_{A, A^{-},H} S$, and let \\
	1. $H = \emptyset$, $A^{-} = A$ : {\em empirical possibility (total)}.		\\
	2. $H = \emptyset$, $A^{-} \neq A$ : {\em empirical possibility (partial)}.		\\
	3. $H \neq \emptyset$, $\Diamond^l_A H$, $\Diamond^e_A H$ : {\em hypothetical possibility} with empirically possible hypothesis.	\\
	4. $H \neq \emptyset$, $\Diamond^l_A H$, $\neg \Diamond^e_A H$, $\neg \Diamond^e_A \neg H$ : {\em hypothetical possibility} with empirically neutral hypothesis.		\\
	5. $H \neq \emptyset$, $\Diamond^l_A H$, $\neg \Diamond^e_A H$, $\Diamond^e_A \neg H$ : {\em hypothetical possibility} with empirically unlikely hypothesis.	\\
	6. $H \neq \emptyset$, $\Diamond^l_A H$, $\Box^e_A \neg H$ : {\em hypothetical possibility} with empirically necessarily false hypothesis.	\\
	7. $H \neq \emptyset$, $\neg \Diamond^l_A H$ : {\em counterfactual possibility}.	\\
\hline
\end{tabular}
\end{table}

As an example, consider four different hypotheses ($H_1$, $H_2$, $H_3$, and $H_4$) with which the same proposition $S$ is hypothetically possible with roughly the same strength (thus, $\Diamond^c_{A, A^{-},H_i} S$ and $\Diamond^l_A H_i$ for every $i=1,...,4$), but the hypotheses are supported by the actual body of evidence $A$ to varying degrees:
\begin{itemize}
\item[1.] $\Diamond^e_A H_1$.
\item[2.] Neither $\Diamond^e_A H_2$ nor $\Diamond^e_A \neg H_2$.
\item[3.] $\Diamond^e_A \neg H_3$, but not $\Diamond^e_A H_3$.
\item[4.] $\Box^e_A \neg H_4$.\footnote{For example, let \\ 
$S$ = ``There are at least three different square formations on Mars, each consisting of six stones of equal size." \\
$H_1$ = ``There are millions of different square formations on Mars, each consisting of four stones of equal size." \\
$H_2$ = ``There are thousands of different square formations on Mars, each consisting of five stones of equal size." \\
$H_3$ = ``There is at least two square formation on Mars, consisting of eight stones of equal size." \\
$H_4$ = ``There is at least one square formation on Mars, consisting of a thousand stones of equal size."}
\end{itemize}
From the first to the fourth case, the hypotheses $H_i$ become increasingly less compatible with the actual body of evidence. While the first and second cases involve hypotheses harmless from the perspective of the actual body of evidence, in the fourth case, the actual body of evidence already compels that the hypothesis does not hold. The gradational scale from the first to the fourth case can thus help evaluate how seriously we should take a certain hypothetical possibility (see Table \ref{table1}). The next section gives further illustrations.

\section{Conditional Inductive Possibility and Counterintuitive Examples of Physical Possibility}\label{section_threeexamples3}

I will illustrate the types and degrees of conditional inductive possibility using my previous examples from Classical Mechanics, General Relativity theory, and Quantum Mechanics.
\begin{itemize}
\item[1.] In the domain of classical physics: \\
(c) The proposition that a stone ball placed on an imagined slope on Mars ($H$) will immediately start rolling down the slope ($S$) is hypothetically possible, provided we draw inductive conclusions based on current evidence supporting Classical Mechanics ($A^{-}$). The hypothesis $H$ stating the relevant initial condition (namely, there is a stone ball on a slope on Mars) is empirically possible, and $A^{-} \cup H$ positively inductively supports that this imagined situation can modeled by Classical Mechanics with an initial condition that does not lead to initial value indeterminism. Thus, $\Diamond^c_{A, A^{-},H} S$, $H \neq \emptyset$, $\Diamond^l_A H$, and $\Diamond^e_A H$. This case corresponds to a hypothetical possibility with the strongest hypothesis. \\
(d) The indeterminism illustrated by Norton's Dome or Xia's point masses ($S$) is hypothetically possible with an empirically necessarily false hypothesis, provided we draw inductive conclusions based on current evidence supporting Classical Mechanics ($A^{-}$). The hypothesis $H$ stating that the relevant initial conditions (i.e., that there exists a point mass positioned at the peak of a Norton's Dome) describe real world situations, for reasons mentioned before, is empirically necessarily false. Thus, $\Diamond^c_{A, A^{-},H} S$, $H \neq \emptyset$, $\Diamond^l_A H$, but $\Box^e_A \neg H$. This case corresponds to a hypothetical possibility with the weakest hypothesis. 
\item[2.] In the domain of relativistic physics: \\
(c) The proposition that after a future Earth-Mars round trip ($H$) the biological age difference between me and my siblings will slightly increase ($S$) is hypothetically possible, provided we draw hypothetical conclusions based on current evidence supporting General Relativity ($A^{-}$). The hypothesis $H$ stating the relevant initial condition (namely, I will participate in such an Earth-Mars round trip in the future) has a different degree of plausibility depending on how near or distant we imagine the trip to be. Thus, $\Diamond^c_{A, A^{-},H} S$, $H \neq \emptyset$, $\Diamond^l_A H$, and if the trip is set for 2025, then $\Box^e_A \neg H$; if the trip is set for 2045, then $\Diamond^e_A H$; and the rest is somewhere between. \\
(d) Time travel of the G\"odel universe ($S$) is trivially counterfactually possible in the sense that if we choose a hypothesis describing the distribution of matter ($H$) which, together with an actual body of evidence that contains Einstein's equation, deductively entails $S$ then, since we essentially condition on $S$, $S$ of course becomes counterfactually possible. The interesting, non-trivial question is whether $S$ could be rendered counterfactually possible by assumptions that do not deductively entail $S$. \\
Analogously to the example of the empty world, it is highly doubtful that there exists a sufficiently narrow $A^{-}$ actual body of evidence supporting General Relativity which is, firstly, sufficiently rich to allow for an application of the material theory of induction, and secondly, does not entail logically that our world cannot be represented by G\"odel's solution (which posits a uniformly rotating matter distribution globally). Lacking an explicit demonstration that such body of evidence $A^{-}$ exists we have good reasons to believe that $S$ is neither hypothetically nor counterfactually possible. \\
(This reasoning does not completely prohibit the counterfactual possibility of time travel, since there are other solutions of Einstein's equation featuring closed timelike curves which, contrary to G\"odel's solution, can be understood as representing only certain parts of our universe.)
\item[3.] In the domain of quantum physics: \\
(c) The proposition that the quantum state of my partner, currently peacefully resting in bed ($H$), two seconds later will evolve into a quantum state where there is a fantastically small probability that, upon looking at my partner, I find them levitating above the bed ($S$) is trivially conditionally inductively possible based on current evidence supporting Quantum Mechanics ($A^{-}$), since $S$ follows deductively from $H$ and the Schr\"odinger equation. \\
(d) The proposition that my partner, currently peacefully resting in bed ($H$), two seconds later will be levitating above the bed when I look at them ($S$) is both hypothetically and counterfactually necessarily false, provided we draw conclusions based on current evidence supporting Quantum Mechanics ($A^{-}$). The probability of levitating as an outcome is fantastically small, so it is inductively compelling that it will never happen. This result follows from the strength of the inductive inference itself and is independent of the empirical possibility of the hypothesis $H$ stating the relevant initial condition.\footnote{See discussion of failure of duality in Section \ref{section_threeexamples2}.}
\end{itemize}

These examples illustrate that Norton's hypothetical and counterfactual modalities are not aligned with physical modalities of the received view. The examples also illustrate how Norton's account provides room for a varied classification of modalities, which, at least in certain cases, lead to conclusions that are more in line with the intuition of a practicing physicist than does the received view.

\section{Summary}\label{section_summary}

\noindent In the first part of this paper, I compared various formulations of physical possibility in the literature, showing that they are special cases of a theory-relative modal formulation of physical possibility. I illustrated through examples from Classical Mechanics, General Relativity, and Quantum Mechanics that this theory-relative formulation of physical possibility can lead to counterintuitive conclusions. To provide further context, I introduced John Norton's concepts of logical, empirical, hypothetical, and counterfactual possibility, along with his material theory of induction.

In his article, Norton claims that physical possibility is empirical possibility. In this paper, I argued that if we understand physical possibility as it is commonly interpreted in the literature, then physical possibility of the received view becomes a special case of Norton's logical possibility, rather than coinciding with his empirical possibility. I also argued that physical possibility of the received view does not coincide with Norton's concepts of hypothetical and counterfactual possibility either. Building on Norton's account, I introduced the notion of conditional inductive possibility and explained that, beyond the strength of the inductive inference itself, different types of conditional inductive possibilities can be identified based on empirical evaluation of the employed hypothesis. I also argued that the introduced gradations help to manage several counterintuitive examples of physical possibility.

Specifically, an empty world is physically possible according to most of our physical theories; uncaused, indeterministic events are physically possible according to Classical Mechanics; G\"odel's time travel is physically possible according to General Relativity; and radical freedom of acting otherwise is physically possible according to Quantum Mechanics. In contrast, an empty world is neither empirically, hypothetically nor counterfactually possible; uncaused, initial value indeterministic events are not empirically possible and are hypothetically possible only under an empirically necessarily false hypothesis; time travel in G\"odel's universe is not empirically possible, and we have good reasons to believe that it is neither hypothetically, nor counterfactually possible; and radical freedom of acting otherwise is neither empirically, hypothetically, nor counterfactually possible.

Thus, contrary to the counterintuitive conclusions of the theory-relative formulation of physical possibility, Norton's concepts of empirical, hypothetical, and counterfactual possibility better align with the intuition of practicing physicists. This alignment provides further support, beyond Norton's own arguments, for considering his inductive modality concepts as capable of capturing the modalities underpinning the practice of scientific theories, particularly physical theories.

\section*{Acknowledgements}

The author would like to thank the audience of the Modality conference at the Research Centre for Humanities for useful questions, and M\'arton G\"om\"ori, John Norton, G\'abor Szab\'o, and referees for useful comments and suggestions on the manuscript. A Hungarian language version of this article is to appear in the {\em Hungarian Philosophical Review}. This research has been supported by the OTKA K-115593 grant.

\theendnotes


\end{document}